\documentclass[twoside,10pt]{article}

\input{jgrg19.sty}

\begin{document}

\pagestyle{fancy}
\fancyhead{}
  \fancyhead[RO,LE]{\thepage}
  \fancyhead[LO]{S.\ A.\ Hughes}                  %% Your name please
  \fancyhead[RE]{Gravitational waves}    %% Your short title please
\rfoot{}
\cfoot{}
\lfoot{}

\title{
  Probing strong-field gravity and black holes\\ with gravitational waves
}

\author{
  S.\ A.\ Hughes\footnote{Email address: sahughes@mit.edu}
}

\address{
  Department of Physics and MIT Kavli Institute, Massachusetts
  Institute of Technology,\\ 77 Massachusetts Avenue, Cambridge, MA,
  02139 United States
}

\abstract{Gravitational wave observations will be excellent tools for
making precise measurements of processes that occur in very
strong-field regions of spacetime.  Extreme mass ratio systems, formed
by the capture of a stellar mass body compact by a massive black hole,
will be targets for planned space-based interferometers such as LISA
and DECIGO.  These systems will be especially powerful tools for
testing the spacetime nature of black hole candidates.  In this
writeup of the talk I gave at JGRG19, I describe how the properties of
black holes are imprinted on their waveforms, and how measurements can
be used to study these properties and thereby learn about the
astrophysics of black holes and about strong-field gravity.}

Detectors for measuring gravitational waves (GWs) have recently
completed their first multiyear data runs.  As this article is
written, some of these detectors are being run at ``enhanced''
sensitivity.  It is expected that a final upgrade to ``advanced''
sensitivity will be needed in order for GWs from astrophysical sources
to be measured regularly.  Once that state is reached, we can turn
this process around, using the GWs that we measure to learn about
their sources, using GWs for observational astronomy.  The purpose of
this article (and the talk on which it is based) is to give a brief
review of the state of this field, focusing in particular on how the
characteristics of black holes and strong-field gravity are imprinted
on a system's GWs.

\section{Gravitational waves: Physics and astrophysics}
\label{gws}

We begin with a brief description of how GWs arise in general
realtivity (GR).  Our purpose is to introduce the main concepts which
describe this phenomenon; later, we will revisit this calculation,
showing how to go to higher order in order to describe realistic
astrophysical sources.  We conclude this section with a quick summary
of the astrophysics of binary GW sources.

\subsection{Leading waveform}
\label{sec:waveform}

We begin by considering ``weak'' gravity, so that spacetime is nearly
that of special relativity,
\begin{equation}
g_{\alpha\beta} = \eta_{\alpha\beta} + h_{\alpha\beta}\;.
\end{equation}
Take the correction to flat spacetime to be small, so that we can
linearize in $h_{\alpha\beta}$ when we build our curvature tensors.
The Einstein tensor in particular becomes
\begin{equation}
G_{\alpha\beta} = \frac{1}{2} \left(\partial_\alpha\partial^\mu
h_{\mu\beta} + \partial_\beta\partial^\mu h_{\mu\alpha} -
\partial_\alpha\partial_\beta h - \Box h_{\alpha\beta} +
\eta_{\alpha\beta}\Box h - \eta_{\alpha\beta}\partial^\mu\partial^\nu
h_{\mu\nu}\right)\;,
\label{eq:lin_einstein1}
\end{equation}
where $h \equiv \eta^{\alpha\beta}h_{\alpha\beta}$ is the trace of
$h_{\alpha\beta}$, and $\Box \equiv \eta^{\alpha\beta}\partial_\alpha
\partial_\beta$ is the flat spacetime wave operator.

Equation (\ref{eq:lin_einstein1}) is rather messy.  To clean it up, we
first introduce the {\it trace-reversed} metric perturbation ${\bar
h}_{\alpha\beta} \equiv h_{\alpha\beta} - (1/2)\eta_{\alpha\beta} h$.
With this definition, Eq.\ (\ref{eq:lin_einstein}) becomes
\begin{equation}
G_{\alpha\beta} = \frac{1}{2} \left(\partial_\alpha\partial^\mu \bar
h_{\mu\beta} + \partial_\beta\partial^\mu \bar h_{\mu\alpha} - \Box
\bar h_{\alpha\beta} - \eta_{\alpha\beta}\partial^\mu\partial^\nu \bar
h_{\mu\nu}\right)\;.
\label{eq:lin_einstein2}
\end{equation}
Next, we take advantage of the {\it gauge-freedom} of linearized
gravity.  In electrodynamics, we may adjust the potential by the
gradient of a scalar, $A_\mu \to A_\mu - \partial_\mu \Lambda$.  This
leaves the field tensor $F_{\mu\nu} = \partial_\mu A_\nu -
\partial_\nu A_\mu$ unchanged.  In linearized GR, a similar operation
follows by adjusting coordinates: If one changes coordinates $x^\alpha
\to x^\alpha + \xi^\alpha$ (requiring $\partial_\mu\xi^\alpha \ll 1$),
then
\begin{equation}
h_{\mu\nu} \to h_{\mu\nu} - \partial_\mu\xi_\nu - \partial_\nu\xi_\mu\;.
\label{eq:gauge_lingrav}
\end{equation}
One can easily show that changing gauge leaves all curvature tensors
unchanged.

We take advantage of our gauge freedom to choose $\xi^\alpha$ so that
$\partial^\mu {\bar h}_{\mu\nu} = 0$.  This is called ``Lorenz gauge''
in analogy with the electrodynamic Lorenz gauge condition
$\partial^\mu A_\mu = 0$.  This simplifies our Einstein tensor
considerably, yielding
\begin{equation}
G_{\alpha\beta} = -\frac{1}{2}\Box{\bar h}_{\alpha\beta}\;.
\label{eq:lin_einstein}
\end{equation}
The Einstein equation for linearized gravity thus takes the simple
form
\begin{equation}
\Box{\bar h}_{\alpha\beta} = -\frac{16\pi G}{c^4} T_{\alpha\beta}\;.
\label{eq:lin_efe}
\end{equation}
Using a radiative Green's function [e.g., \cite{jackson}, Sec.\
12.11], we find the solution
\begin{equation}
{\bar h}_{\alpha\beta}({\bf x}, t) =
    \frac{4G}{c^4}\int\frac{T_{\alpha\beta}({\bf x}', t - |{\bf x} -
    {\bf x}'|/c)}{|{\bf x} - {\bf x}'|} d^3x'\;.
\label{eq:lin_soln}
\end{equation}
Here, ${\bf x}$ is a spatial ``field point,'' where $\bar
h_{\alpha\beta}$ is evaluated, and ${\bf x}'$ is a ``source point,''
the spatial coordinate we integrate over the source's extent.  Notice
that the solution at $t$ depends on what happens to the source at {\it
retarded time} $t - |{\bf x} - {\bf x'}|/c$.  Information must
causally propagate from ${\bf x'}$ to ${\bf x}$.

Equation (\ref{eq:lin_soln}) is an exact solution to the linearized
field equation.  It gives the unfortunate impression that every
component of the metric perturbation is radiative.  Just as one can
choose a gauge such that an isolated point charge has an oscillatory
potential, the Lorenz gauge makes {\it all} components of the metric
appear radiative, even if they are static\footnote{In the
electromagnetic case, it is unambiguous which {\it field} components
are radiative and which are static.  Similarly, one can always tell
which {\it curvature} components are radiative and which are static.
Eddington {\cite{eddington22}} appears to have been the first to use
the curvature tensor to categorize gravitational degrees of freedom in
this way.}.  Fortunately, it is not difficult to see that only a
subset of the metric represents the radiative degrees of freedom in
{\it all} gauges.  We will only quote the result here; interested
readers can find the full calculation in Ref.\ \cite{njp}, Sec.\ 2.2:
{\it Given a solution $h_{\alpha\beta}$ to the linearized Einstein
field equations, only the {\bf spatial}, {\bf transverse}, and {\bf
traceless} components $h^{\rm TT}_{ij}$ describe the spacetime's
gravitational radiation in a gauge-invariant manner.}  Traceless means
\begin{equation}
\delta_{ij} h^{\rm TT}_{ij} = 0\;;
\label{eq:traceless}
\end{equation}
``transverse'' means
\begin{equation}
\partial_i h^{\rm TT}_{ij} = 0\;.
\label{eq:transverse}
\end{equation}
Expanding $h^{\rm TT}_{ij}$ in Fourier modes, we see that Eq.\
(\ref{eq:transverse}) requires $h^{\rm TT}_{ij}$ to be orthogonal (in
space) to each mode's wave vector ${\bf k}$.

Conditions (\ref{eq:traceless}) and (\ref{eq:transverse}) make it
simple to construct $h^{\rm TT}_{ij}$ given some $h_{ij}$.  Let $n_i$
denote components of the unit vector along the propagation direction.
The tensor
\begin{equation}
P_{ij} = \delta_{ij} - n_in_j
\end{equation}
projects spatial components orthogonal to ${\bf n}$.  It is then
simple to verify that
\begin{equation}
h^{\rm TT}_{ij} = h_{kl}\left(P_{ki}P_{lj} -
\frac{1}{2}P_{kl}P_{ij}\right)
\label{eq:projected_hTT}
\end{equation}
represents the ``TT'' metric perturbation.  We can now manipulate the
solution (\ref{eq:lin_soln}) into
\begin{equation}
h^{\rm TT}_{ij} =
\frac{2}{D}\frac{G}{c^4}\frac{d^2I_{kl}}{dt^2}\left(P_{ik}P_{jl} -
\frac{1}{2}P_{kl}P_{ij}\right)\;,
\label{eq:quadrupole1}
\end{equation}
where $D$ is distance to the source, and where
\begin{equation}
I_{ij} = \int x^{i'}x^{j'} T_{tt}({\bf x}', t)\,d^3x'
\end{equation}
is the source's {\it quadrupole moment}.  It is straightforward to
show that the trace $I \equiv I_{ii}$ does not contribute to Eq.\
(\ref{eq:quadrupole1}), so it is common to use the ``reduced''
quadrupole moment,
\begin{equation}
{\cal I}_{ij} = I_{ij} - \frac{1}{3}\delta_{ij}I\;.
\end{equation}
The waveform then takes the form in which it is usually presented,
\begin{equation}
h^{\rm TT}_{ij} =
\frac{2}{R}\frac{G}{c^4}\frac{d^2{\cal I}_{kl}}{dt^2}\left(P_{ik}P_{jl} -
\frac{1}{2}P_{kl}P_{ij}\right)\;,
\label{eq:quadrupole}
\end{equation}
the {\it quadrupole formula} for GW emission.

GWs also carry energy from their source.  Isaacson {\cite{isaacson68}}
first carefully analyzed this in a tensorial manner, showing that GWs
produce a stress-energy tensor given by
\begin{equation}
T_{\mu\nu}^{\rm GW} = \frac{c^4}{32\pi G} \langle \hat\nabla_\mu
h_{\alpha\beta} \hat\nabla_\nu h^{\alpha\beta} \rangle\;,
\label{eq:isaacson_tmunu}
\end{equation}
where $\hat\nabla_\mu$ denotes a covariant derivative in the
background spacetime.  (This assumes the waveform is in a gauge such
that it is transverse and traceless; more general expressions exist.)
Notice that the energy content is quadratic in the wave amplitude;
computing it correctly requires taking our perturbative analysis to
second order.  We defer the details of this derivation to Ref.\
{\cite{isaacson68}}.

Now consider a binary system with Newtonian orbital dynamics,
radiating GWs according to Eq.\ (\ref{eq:quadrupole}) and evolving by
energy and angular momentum carried off in accordance with Eq.\
(\ref{eq:isaacson_tmunu}).  Begin with the binary's members in
circular orbit of separation $R$.  This binary has orbital energy
\begin{equation}
E^{\rm orb} = \frac{1}{2}m_1 v_1^2 + \frac{1}{2}m_2 v_2^2 -
\frac{Gm_1m_2}{R} = -\frac{G\mu M}{2R}\;,
\end{equation}
(where $M = m_1 + m_2$ and $\mu = m_1 m_2/M$) and orbital frequency
\begin{equation}
\Omega_{\rm orb} = \sqrt{\frac{GM}{R^3}}\;.
\end{equation}
Far from the source, Eq.\ (\ref{eq:isaacson_tmunu}) tells us the flux
of energy carried by GWs:
\begin{equation}
\frac{dE^{\rm GW}}{dAdt} = \frac{c^4}{32\pi G}\langle \partial_t
h^{\rm TT}_{ij} \partial_k h^{\rm TT}_{ij}\rangle n^k\;.
\end{equation}
Plugging in Eq.\ (\ref{eq:quadrupole}) and integrating over the
sphere, we find
\begin{equation}
\frac{dE}{dt}^{\rm GW} = \int dA\,\frac{dE}{dAdt} =
\frac{G}{5c^5}\left\langle \frac{d^3{\cal I}_{ij}}{dt^3} \frac{d^3{\cal
I}_{ij}}{dt^3}\right\rangle\;.
\label{eq:quadrupole_edot}
\end{equation}
For the Newtonian binary,
\begin{equation}
{\cal I}_{ij} = \mu \left(x_i x_j -
\frac{1}{3}R^2\delta_{ij}\right)\;;
\end{equation}
we choose coordinates such that the components of the separation
vector are $x_1 = R\cos\Omega_{\rm orb} t$, $x_2 = R\sin\Omega_{\rm
orb} t$, $x_3 = 0$.  Inserting into Eq.\ (\ref{eq:quadrupole_edot}),
we find
\begin{equation}
\frac{dE}{dt}^{\rm GW} = \frac{32}{5}\frac{G}{c^5} \mu^2 R^4
\Omega^6\;.
\label{eq:circbin_edot}
\end{equation}
We now assert that the binary evolves quasi-statically --- any
radiation carried off by GWs is accounted for by the evolution of its
orbital energy, $dE^{\rm orb}/dt + dE^{\rm GW}/dt = 0$.  Allow the
orbital radius to slowly change in time, so that $dE^{\rm orb}/dt =
(dE^{\rm orb}/dR)(dR/dt)$.  Combining this rule with Eq.\
(\ref{eq:circbin_edot}), we find
\begin{equation}
R(t) = \left[\frac{256 G^3 \mu M^2(t_c - t)}{5c^5}\right]^{1/4}\;.
\label{eq:rorb_of_time}
\end{equation}
This in turn tells us that the orbital frequency changes according to
\begin{equation}
\Omega_{\rm orb}(t) = \left[\frac{5c^5}{256(G{\cal M})^{5/3}(t_c -
t)}\right]^{3/8}\;.
\label{eq:Newt_quad}
\end{equation}
We have introduced the {\it chirp mass} ${\cal M} \equiv
\mu^{3/5}M^{2/5}$, so called because it sets the rate at which the
binary sweeps upward in frequency, or ``chirps.''  We have also
introduced the coalescence time $t_c$, which formally describes when
the separation goes to zero (or when frequency goes to infinity).
Corrections for eccentricity can be computed by separately accounting
for the evolution of the binary's energy and angular momentum; see
Ref.\ {\cite{shapteuk}}, Exercise 16.10 for details.

We conclude this section by writing the gravitational waveform
predicted for quadrupole emission from the Newtonian, circular binary.
Evaluating Eq.\ (\ref{eq:quadrupole}), we find that $h_{ij}$ has two
polarizations.  These are labeled ``plus'' and ``cross,'' from the
lines of force associated with their tidal stretch and squeeze:
\begin{eqnarray}
h_+ &=& -\frac{2G{\cal M}}{c^2D}\left(\frac{\pi G{\cal
M}f}{c^3}\right)^{2/3}(1 + \cos^2\iota) \cos2\Phi_N(t)\;,
\nonumber\\
h_\times &=& -\frac{4G{\cal M}}{c^2D}\left(\frac{\pi G{\cal
M}f}{c^3}\right)^{2/3}\cos\iota \sin2\Phi_N(t)\;,
\label{eq:h_NQ}
\end{eqnarray}
where the phase
\begin{equation}
\Phi_N(t) = \int \Omega_{\rm orb}\,dt = \Phi_c - \left[\frac{c^3(t_c
- t)}{5G{\cal M}}\right]^{5/8}\;,
\label{eq:phi_NQ}
\end{equation}
and where $f = (1/\pi)d\Phi_N/dt$ is the GW frequency.  The system's
inclination $\iota$ is just the projection of its orbital angular
momentum, ${\bf L}$, to the wave's direction of propagation ${\bf n}$:
$\cos\iota = \hat{\bf L}\cdot{\bf n}$ (where $\hat{\bf L} = {\bf
L}/|{\bf L}|$).  Note that $h_+$ and $h_\times$ depend on, and thus
encode, the chirp mass, distance, the position on the sky (via the
direction vector ${\bf n}$), and the orientation of the binary's
orbital plane (via $\hat{\bf L}$).  In later discussion, we will amend
Eq.\ (\ref{eq:h_NQ}) and (\ref{eq:phi_NQ}) to include higher order
contributions to the binary's waves and evolution.

\subsection{Astrophysical binary sources}
\label{sec:astro}

The binary example considered in the previous section is particularly
germane since compact binary systems are among the most important
astrophysical sources of GWs.  Indeed, our best data on GWs and GW
sources comes from observations of {\it binary pulsar} systems, pairs
of neutron stars at least one of which is a pulsar.  The pulsar member
of the pair acts as an outstanding clock, allowing the properties of
the binary to be mapped with great precision.

Some binary neutron stars are in such strong field orbits that the
evolution of the binary's orbital period due to GW emission can be
discerned over long observational baselines.  The prototypical example
is the first such system discovered, PSR 1913+16.  Over 30 years of
study have found extraordinary agreement between prediction and
observation for the evolution of this system's orbit {\cite{wt05}}.
Additional inspiraling systems have been discovered; in all cases for
which we have enough data to discern period evolution, the data agree
with theory to within measurement precision {\cite{stairs98, nice05,
jacoby06, kramerstairs08, bhat08}}.  At least one additional recently
discovered system is likely to show a measurable inspiral in the next
few years {\cite{kasian08}}.

Turn from binary neutron stars to compact binaries more generally.
Such systems are organized most naturally by their masses.  At the low
end we have {\it stellar-mass} binaries, including binary pulsars.
The data on these binaries are quite solid, since we can tie models
for their birth and evolution to observations.  At least some fraction
of short gamma-ray bursts are likely to be associated with the mergers
of neutron star-neutron star (NS-NS) or black hole-neutron star
(BH-NS) systems {\cite{eichler89, fox05}}; as such, gamma-ray
telescopes may already be telling us about compact binary merger many
times per year {\cite{nakar06}}.

There is also evidence that nature produces {\it supermassive}
binaries, in which the members are black holes with $M \sim 10^6 -
10^8\,M_\odot$ such as are found at the centers of galaxies.
Theoretical arguments combining hierarchical galaxy growth scenarios
with the hypothesis that most galaxies host black holes generically
predict the formation of such binaries.  We have now identified many
systems with properties indicating that they may host such binaries.
The evidence includes active galaxies with double cores
{\cite{komossa03, maness04, rodriguez06}}; systems with doubly-peaked
emission lines {\cite{zhou04, gerke07}}; helical radio jets
{\cite{bbr80, cw95, lr05}}; and periodic or semi-periodic systems,
such as the blazar OJ287 {\cite{valtonen08}}.  As surveys go deeper
and resolution improves, we may expect the catalog of candidate
supermassive black hole binaries to expand.

Now consider theoretical models.  Assuming that our galaxy is typical
and that the inferred density of NS-NS systems in the Milky Way
carries over to similar galaxies (correcting for factors such as
typical stellar age and the proportion of stars that form neutron
stars), we can estimate the rate at which binary systems merge in the
universe.  References {\cite{nps91}} and {\cite{phinney91}} first made
such estimates, finding a ``middle-of-the-road'' rate that about 3
binaries per year merge to a distance of 200 Mpc.  More recent
calculations based on later surveys and observations of NS-NS systems
have amended this number somewhat; the total number expected to be
measured by advanced detectors is around several tens per year.  See,
for example, {\cite{kalogera07}} for a detailed discussion of
methodology.

Another technique uses population synthesis.  These calculations
combine data on the observed distribution of stellar binaries with
models for how stars evolve.  This allows us to estimate the rate of
formation and merger for systems which we cannot at present observe,
such as stellar mass black hole-black hole (BH-BH) binaries, or for
which we have only circumstantial evidence, such as neutron star-black
hole (NS-BH) binaries (which presumably form some fraction of short
gamma ray bursts).  A disadvantage is that the models of stellar
evolution in binaries have many uncertainties.  There are multiple
branch points in binary evolution, such as whether the binary remains
bound following each supernova, and whether the binary survives common
envelope evolution.  As a consequence, the population synthesis
predictions can be quite diverse.  Though different groups generally
agree well with the rates for NS-NS systems (by design), their
predictions for NS-BH and BH-BH systems differ by quite a bit.  New
data are needed to clear the theoretical cobwebs.

Binaries can also form dynamically in dense environments, such as
globular clusters.  The most massive bodies will tend to sink to a
cluster's core through mass segregation {\cite{spitzer69}}; as such,
the core will become populated with the heaviest bodies, either stars
which will evolve into compact objects, or the compact objects
themselves.  As those objects interact with one another, they will
tend to form massive binaries; calculations show that the production
of BH-BH binaries is particularly favored.  It is thus likely that
globular clusters will act as ``engines'' for the production of
massive compact binaries {\cite{pzm00, oor07, mwdg08}}.

The hierarchical growth scenario for galaxies, coupled with the
hypothesis that most galactic bulges host large black holes,
generically predicts the formation of supermassive binaries,
especially at high redshifts when mergers were common.  The first
careful discussion of this was by Begelman, Blandford, and Rees
{\cite{bbr80}}.  The coevolution of black holes and galaxies in
hierarchical scenarios has now become a very active focus of research
(e.g., Refs.\ {\cite{mhn01, yt02, vhm03}}).  Galaxy mergers appear to
be a natural mechanism to bring ``fuel'' to one or both black holes,
igniting quasar activity; the formation of a binary may thus be
associated with the duty cycle of quasars {\cite{hco04, hckh08,
dcshs08}}.  Such scenarios typically find that most black hole mergers
come at fairly high redshift ($z \gtrsim 3$ or so), and that the bulk
of a given black hole's mass is due to gas it has accreted over its
growth.

A subset of binaries in the supermassive range are of particular
interest to the relativity theorist.  These binaries form by the
capture of a ``small'' ($1 - 100\,M_\odot$) compact object onto an
orbit around a black hole in a galactic center.  Such binaries form
dynamically through stellar interactions {\cite{sr97, ha05}}; the
formation rate predicted by most models is typically $\sim 10^{-7}$
extreme mass ratio binaries per galaxy per year \cite{ha05}.  If the
inspiraling object is a white dwarf or star, it could tidally disrupt
as it comes close to the massive black hole, producing an x-ray or
gamma-ray flare {\cite{rbb05, mhk08}}.  If the inspiraling object is a
neutron star or black hole, it will be swallowed whole by the large
black hole.  As such, it will almost certainly be electromagnetically
quiet; however, its GW signature will be loud, and is a particularly
interesting target.

\vfill

\section{Measuring gravitational waves: Principles and experiments}
\label{sec:measure}

Before moving to a discussion of how black hole characteristics and
strong-field physics are imprinted on GWs, let us briefly summarize
the key principles by which a GW interferometer operators.  Begin with
the simple limit in which we treat the spacetime in which our detector
lives as flat plus a simple GW propagating down our coordinate
system's $z$-axis:
\begin{equation}
ds^2 = -c^2dt^2 + (1 + h)dx^2 + (1 - h)dy^2 + dz^2\;,
\label{eq:detector_spacetime}
\end{equation}
where $h = h(t - z)$.  We neglect the influence of the earth (clearly
important for terrestrial experiments) and the solar system (which
dominates the spacetime of space-based detectors).  Corrections
describing these influences can be added; we neglect them as they vary
on much longer timescales than the GWs.

\begin{figure}[t]
\begin{center}
\includegraphics[width=4in]{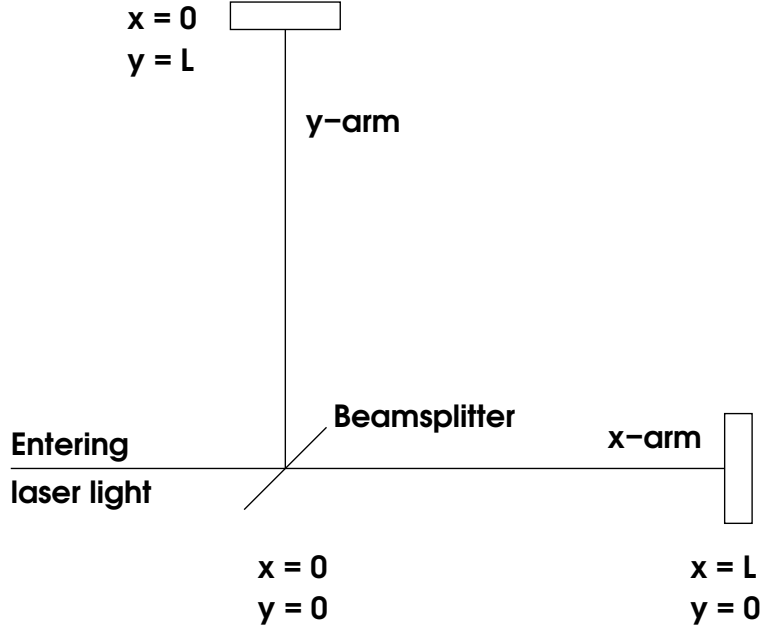}
\end{center}
\caption{Schematic of an interferometer that could be used to detect
GWs.  Though real interferometers are vastly more complicated, this
interferometer topology contains enough detail to illustrate the
principle by which such measurements are made.}
\label{fig:interf}
\end{figure}

Figure {\ref{fig:interf}} sketches an interferometer that can measure
a GW.  Begin by examining the geodesics describing the masses at the
ends of the arms, and the beam splitter at the center.  Take these
objects to be initially at rest, so that $(dx^\mu/d\tau)_{\rm before}
\doteq (c,0,0,0)$.  The GW shifts this velocity by an amount of order
the wave strain: $(dx^\mu/d\tau)_{\rm after} = (dx^\mu/d\tau)_{\rm
before} + {\cal O}(h)$.  Now examine the geodesic equation:
\begin{equation}
\frac{d^2x^j}{d\tau^2} + {\Gamma^j}_{\alpha\beta}
\frac{dx^\alpha}{d\tau}\frac{dx^\beta}{d\tau} = 0\;.
\end{equation}
All components of the connection are ${\cal O}(h)$.  Combining this
with our argument for how the GW affects the various velocities, we
have
\begin{equation}
\frac{d^2x^j}{d\tau^2} + {\Gamma^j}_{00}
\frac{dx^0}{d\tau}\frac{dx^0}{d\tau} + {\cal O}(h^2) = 0\;.
\end{equation}
It is simple to show that the connection coefficient $ {\Gamma^j}_{00}
= 0$, as the relevant metric components are all constant.  We conclude
that
\begin{equation}
\frac{d^2x^j}{d\tau^2} = 0 \;.
\end{equation}
In other words, {\it the test masses are unaccelerated to leading
order in the GW amplitude $h$.}

This seems to say that the GW has no impact!  However, the geodesic
equation describes motion {\it with respect to specified coordinates}.
Our coordinates are effectively ``comoving'' with the interferometer's
components.  Using the fact that our mirrors are at constant position
in these coordinates, it is simple to see that the {\it proper} length
of the arms does change.  For instance, the $x$-arm has a proper
length
\begin{equation}
D_x = \int_0^L \sqrt{g_{xx}}\,dx = \int_0^L \sqrt{1 + h}\,dx \simeq
\int_0^L \left(1 + \frac{h}{2}\right)dx = L\left(1 +
\frac{h}{2}\right)\;.
\label{eq:proper_length}
\end{equation}
Likewise, the $y$-arm has a proper length $D_y = L(1 - h/2)$.

This means that the armlengths as measured by a ruler will vary with
$h$.  One might worry that, in practice, the ruler will vary with the
wave, cancelling the measurement.  This does not happen because rulers
are not made of freely-falling particles: Its elements are {\it bound}
to one another, and act against the GW.  The ruler feels some effect
due to the GW, but it is far smaller than the variation in $D_x$ and
$D_y$.

The ruler used by the most sensitive current and planned detectors is
based on laser interferometry.  We will not describe the details of
how a GW is imprinted on the output observable of an interferometer
such as that sketch in Fig.\ {\ref{fig:interf}}; for our purposes, it
is enough to note that in essence one uses the (highly stable)
frequency of the laser as a clock, and times the light travel in the
two arms.  We recommend the nicely pedagogical article by Faraoni
{\cite{faraoni07}} for a clear discussion, as well as a relatively
recent analysis by Finn {\cite{finn08}} for more detailed discussion.

From basic principles, we now give a brief summary of current and
planned detectors.  Our goal is not an in-depth discussion, so we
refer readers interested in these details to excellent reviews by
\cite{hr00} (which covers in detail the characteristics of the various
detectors) and \cite{td05} (which covers the interferometry used for
space-based detectors).  When thinking about GW detectors, a key
characteristic is that the frequency of peak sensitivity scales
inversely with armlength.  The ground-based detectors currently in
operation are sensitive to waves oscillating at 10s -- 1000s of Hertz.
Planned space-based detectors will have sensitivities at much lower
frequencies, ranging from $10^{-4}$ -- 0.1 Hz (corresponding to waves
with periods of tens of seconds to hours).

The ground-based detectors in operation are LIGO ({\it Laser
Interferometer Gravitational-wave Observatory}), with antennae in
Hanford, Washington and Livingston, Louisiana; Virgo near Pisa, Italy;
and GEO near Hanover, Germany.  The LIGO interferometers have
4-kilometer arms, and a peak sensitivity near 100 Hz.  Virgo has
3-kilometer arms, and sensitivity comparable to the LIGO detectors.
GEO has 600-meter arms; as such, its peak sensitivity is at higher
frequencies than LIGO and Virgo.  Using advanced interferometry
techniques, it achieves sensitivity competitive with the
kilometer-scale instruments.  All of these instruments will be
upgraded over the course of the next few years, installing more
powerful lasers, and reducing the impact of local ground vibrations.
The senstivity of LIGO should be improved by roughly a factor of ten,
and the bandwidth increased as well.  See {\cite{fritschel03}} for
detailed discussion.

There are plans to build additional kilometer-scale instruments.  The
detector AIGO ({\it Australian International Gravitational
Observatory}) is planned as a detector very similar to LIGO and Virgo,
but in Western Australia {\cite{mcc02}}.  This location, far from the
other major GW observatories, has great potential to improve the
ability of the worldwide GW detector network to determine the
characteristics of GW events {\cite{ssmf06}}.  The Japanese GW
community, building on their experience with the 300-meter TAMA
interferometer, hopes to build a 3-kilometer {\it underground}
instrument.  Dubbed LCGT ({\it Large-scale Cryogenic
Gravitational-wave Telescope}), the underground location takes
advantage of the fact that local ground motions tend to decay fairly
rapidly as we move away from the earth's surface.  They plan to use
cryogenic cooling to reduce noise from thermal vibrations.

\begin{figure}[t]
\begin{center}
\includegraphics[width=4in]{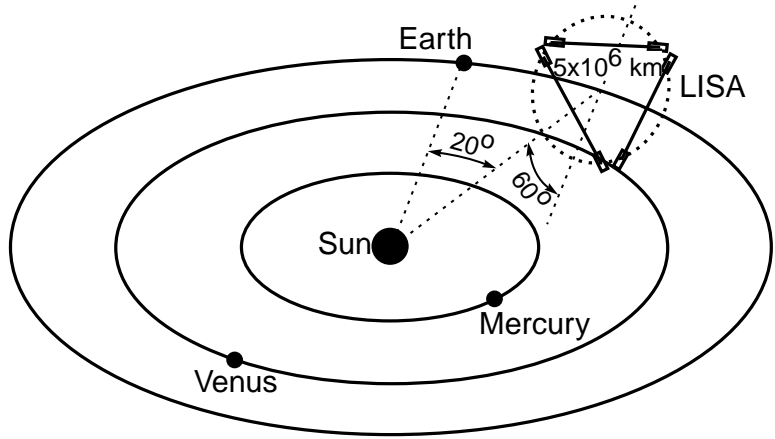}
\end{center}
\caption{Schematic of the LISA constellation in orbit about the sun.
Each arm of the triangle is $5\times10^6$ km; the centroid of the
constellation lags the Earth by $20^\circ$, and its plane is inclined
to the ecliptic by $60^\circ$.  Note that the spacecraft orbit freely;
there is no formation flying in the LISA configuration.  Instead, each
spacecraft is in a slightly eccentric, slightly inclined orbit; their
individual motions preserve the near-equilateral triangle pattern with
high accuracy for a timescale of decades.}
\label{fig:lisa_orb}
\end{figure}

In space, the major project is LISA ({\it Laser Interferometer Space
Antenna}), a 5-million kilometer interferometer under development as a
joint NASA-ESA mission.  LISA will consist of three spacecraft placed
in orbits so that their relative positions form an equilateral
triangle lagging the earth by $20^\circ$, inclined to the ecliptic by
$60^\circ$; see Fig.\ {\ref{fig:lisa_orb}}.  The spacecraft are free
and do not maintain this constellation precisely; however, their
armlength variations occur on a timescale far longer than the periods
of their target waves.  The review by {\cite{td05}} discusses in great
detail how one does interferometry on such a baseline with
time-changing armlengths.  LISA targets waves with periods of hours to
several seconds, a rich band for signals involving massive black
holes.  The LISA {\it Pathfinder}, a testbed for some of the mission's
components, is scheduled for launch in the very near future
{\cite{vitale05}}.

Somewhat smaller than LISA, The Japanese GW community has proposed
DECIGO ({\it DECI-hertz Gravitational-wave Observatory}), a space
antenna to target a band at roughly $0.1$ Hz.  This straddles the peak
sensitivities of LISA and terrestrial detectors, and may thus act as a
bridge for signals that evolve from one band to the other.  See Ref.\
{\cite{decigo}} for further discussion.

\section{Comparable mass binary waves}
\label{sec:comparable}

We now at last begin to examine how the characteristics of black holes
and strong-field gravity are imprinted on the GWs these systems
generate.  We first must go somewhat beyond the leading-order waveform
discussed in Sec.\ {\ref{sec:waveform}}.  After developing the
necessary formal tools, we discuss how the interesting characteristics
appear in the waves.

\subsection{Going beyond leading order}
\label{sec:pn_formal}

In the analytic treatment of comparable mass binary waves, one begins
by considering the {\it post-Newtonian}, or pN, expansion.  This
expansion in turn begins by considering the binary in so-called
harmonic or deDonder coordinates.  In these coordinates, one defines
\begin{equation}
h^{\mu\nu} \equiv \sqrt{-g}g^{\mu\nu} - \eta^{\mu\nu}\;,
\label{eq:h_harmonic}
\end{equation}
where $g$ is the determinant of $g_{\mu\nu}$.  This looks similar to
the flat spacetime perturbation defined in Sec.\ {\ref{sec:waveform}};
however, we do not assume that $h$ is small.  We next impose the gauge
condition
\begin{equation}
\partial_\alpha h^{\alpha\beta} = 0\;.
\end{equation}
With these definitions, the {\it exact} Einstein field equations are
\begin{equation}
\Box h^{\alpha\beta} = \frac{16\pi G}{c^4}\tau^{\alpha\beta}\;,
\label{eq:pn_efe}
\end{equation}
where $\Box = \eta^{\alpha\beta}\partial_\alpha\partial_\beta$ is the
{\it flat} spacetime wave operator.  The form of Eq.\
(\ref{eq:pn_efe}) means that the radiative Green's function we used to
derive Eq.\ (\ref{eq:lin_soln}) can be applied here, yielding
\begin{equation}
h^{\alpha\beta} = -\frac{4G}{c^4}\int \frac{\tau_{\alpha\beta}({\bf x}',
t - |{\bf x} - {\bf x}'|/c)}{|{\bf x} - {\bf x}'|}d^3x'\;.
\label{eq:pn_formal_soln}
\end{equation}
Equation (\ref{eq:pn_formal_soln}) is exact.  Note, however, that we
never defined the source $\tau^{\alpha\beta}$.  It is given by
\begin{equation}
\tau^{\alpha\beta} = (-g)T^{\alpha\beta} +
\frac{c^4\Lambda^{\alpha\beta}}{16\pi G}\;;
\label{eq:pn_tau_def}
\end{equation}
$T^{\alpha\beta}$ is the usual stress energy tensor,
$\Lambda^{\alpha\beta}$ encodes the nonlinear structure of the
Einstein field equations:
\begin{eqnarray}
\Lambda^{\alpha\beta} &\equiv& 16\pi(-g)t^{\alpha\beta}_{\rm LL} +
\partial_\nu h^{\alpha\mu}\partial_\mu h^{\beta\nu} -
\partial_\mu \partial_\nu h^{\alpha\beta} h^{\mu\nu}
\\
&=& N^{\alpha\beta}[h,h] + M^{\alpha\beta}[h,h,h]
+ L^{\alpha\beta}[h,h,h,h] + {\cal O}(h^5)\;.
\end{eqnarray}
On the first line, $t^{\alpha\beta}_{\rm LL}$ is the Landau-Lifshitz
pseudotensor, a quantity which (in certain gauges) allows us to
describe how GWs carry energy through spacetime (\cite{ll75}, Sec.\
96).  On the second line, the term $N^{\alpha\beta}[h,h]$ means a
collection of terms quadratic in $h$ and its derivatives,
$M^{\alpha\beta}[h,h,h]$ is a cubic term, etc.  Our solution
$h^{\alpha\beta}$ appears on both the left- and right-hand sides of
Eq.\ (\ref{eq:pn_formal_soln}).  Such a structure can be handled very
well {\it iteratively}.  We write
\begin{equation}
h^{\alpha\beta} = \sum_{n = 1}^\infty G^n h_n^{\alpha\beta}\;.
\label{eq:iteration}
\end{equation}
The $n = 1$ term is essentially the linearized solution from Sec.\
{\ref{sec:waveform}}.  To go higher, let $\Lambda_n^{\alpha\beta}$
denote the contribution of $\Lambda^{\alpha\beta}$ to the solution
$h_n^{\alpha\beta}$.  We find
\begin{equation}
\Lambda_2^{\alpha\beta} = N^{\alpha\beta}[h_1,h_1]\;,
\end{equation}
\begin{equation}
\Lambda_3^{\alpha\beta} = M^{\alpha\beta}[h_1,h_1,h_1] +
N^{\alpha\beta}[h_2,h_1] + N^{\alpha\beta}[h_1,h_2]\;,
\end{equation}
etc.; higher contributions to $\Lambda^{ab}$ can be found by expanding
its definition and gathering terms.  By solving the equations which
result from this procedure, we can build the spacetime metric and
describe the motion of the members of a binary and the radiation that
they emit.

We defer details of this construction to the literature (Blanchet's
review, Ref.\ {\cite{blanchet06}}, is particularly useful for this
purpose), and turn to a study of the interesting features of the pN
binary waveform.  Take the members of the binary to have masses $m_1$
and $m_2$, let their separation be $r$, and let $\mathbf{\hat r}$
point to body 1 from body 2.  In the harmonic gauge used for pN
theory, the acceleration of body 1 is given by
\begin{equation}
{\bf a} = {\bf a}_0 + {\bf a}_2 + {\bf a}_4 + {\bf a}_5 + {\bf a}_6 +
{\bf a}_7 \ldots \;.
\label{eq:pNorbitaccel}
\end{equation}
The zeroth term,
\begin{equation}
{\bf a}_0 = -\frac{G m_2}{r^2} \mathbf{\hat r},
\end{equation}
is the usual Newtonian gravitational acceleration.  Each ${\bf a}_n$
is a correction of order $(v/c)^n$.  The first is
\begin{equation}
{\bf a}_2 = \left[\frac{5G^2m_1m_2}{r^3} + \frac{4G^2m_2^2}{r^3} +
\frac{Gm_2}{r^2} \left(\frac{3}{2}({\mathbf{\hat r}}\cdot{\bf v_2})^2
- v_1^2 + 4{\bf v_1}\cdot{\bf v_2} -
2v_2^2\right)\right]\frac{\mathbf{\hat r}}{c^2}\;.
\label{eq:pN_a2}
\end{equation}
For the acceleration of body 2 due to body 1, exchange labels 1 and 2
and replace $\mathbf{\hat r}$ with $-\mathbf{\hat r}$.  So far, the pN
acceleration has been computed to order $(v/c)^7$.  As we go to high
order, the expressions for ${\bf a}_n$ become quite lengthy.  An
excellent summary is given in Ref.\ {\cite{blanchet06}}, Eq.\ (131)
and surrounding text.

PN theory also introduces a distinctly non-Newtonian element to binary
dynamics: its members' spins {\it precess} in the binary's curved
spacetime.  If the spins are ${\bf S}_1$ and ${\bf S}_2$, one finds
{\cite{th85}}
\begin{equation}
\frac{d{\bf S}_1}{dt} = \frac{G}{c^2r^3}\left[\left(2 +
\frac{3}{2}\frac{m_2}{m_1}\right)\mu\sqrt{M r}\hat{\bf L}\right]
\times{\bf S}_1 + \frac{G}{c^2r^3}\left[\frac{1}{2}{\bf S}_2 -
\frac{3}{2}({\bf S}_2\cdot\hat{\bf L})\hat{\bf L}\right] \times{\bf
S}_1\;,
\label{eq:dS1dt}
\end{equation}
\begin{equation}
\frac{d{\bf S}_2}{dt} = \frac{G}{c^2r^3}\left[\left(2 +
\frac{3}{2}\frac{m_1}{m_2}\right)\mu\sqrt{M r}\hat{\bf L}\right]
\times{\bf S}_2 + \frac{G}{c^2r^3}\left[\frac{1}{2}{\bf S}_1 -
\frac{3}{2}({\bf S}_1\cdot\hat{\bf L})\hat{\bf L}\right] \times{\bf
S}_2\;.
\label{eq:dS2dt}
\end{equation}
We now discuss the ways in which aspects of pN binary dynamics color a
system's waves.

\subsubsection{Gravitational-wave amplitudes.}
\label{sec:pn_amplitude}

Although a binary's {\it dominant} waves come from variations in its
mass quadrupole moment, higher moments also generate GWs.  In the pN
framework, these moments contribute to the amplitude of a binary's
waves beyond the quadrupole form, Eq.\ (\ref{eq:h_NQ}).  Write the
gravitational waveform from a source as
\begin{equation}
h_{+,\times} = \frac{2G{\cal M}}{c^2D}\left(\frac{\pi G{\cal
M}f}{c^3}\right)^{2/3} \left[H^0_{+,\times} +
v^{1/2}H^{1/2}_{+,\times} + v H^1_{+,\times} + \ldots\right] \;,
\label{eq:hpn_sum}
\end{equation}
where $v \equiv (\pi G M f/c^3)^{1/3}$ is roughly the orbital speed of
the binary's members (normalized to $c$).  The $H^0_{+,\times}$ terms
reproduce the waveform presented in Eq.\ (\ref{eq:h_NQ}).  The
higher-order terms $H^{1/2}_{+,\times}$ and $H^1_{+,\times}$ can be
found in \cite{blanchet06}, his Eqs.\ (237) through (241).  A key
point to note is that these higher-order terms introduce new
dependences on the binary's orbital inclination and its masses.  As
such, measurement of these terms provides additional constraints on
the system's characteristics.

\subsubsection{Orbital phase.}
\label{sec:pn_phase}

The motion of a binary's members about each other determines the
orbital phase.  Specializing to circular orbits, we can determine the
orbital frequency from the acceleration of the its members;
integrating up this frequency, we define the phase $\Phi(t)$.  The
first few terms of this phase are given by {\cite{bdiww95}}
\begin{eqnarray}
\Phi &=& \Phi_c - \left[\frac{c^3(t_c - t)}{5G{\cal M}}\right]^{5/8}
\left[1 + \left(\frac{3715}{8064} +
\frac{55}{96}\frac{\mu}{M}\right)\Theta^{-1/4} -\frac{3}{16}\left[4\pi
- \beta(t)\right]\Theta^{-3/8} \right.  \nonumber\\ & & \left.+
\left(\frac{9275495}{14450688} + \frac{284875}{258048}\frac{\mu}{M} +
\frac{1855}{2048}\frac{\mu^2}{M^2} + \frac{15}{64}\sigma(t)\right)
\Theta^{-1/2}\right]\;,
\label{eq:2pnPhase}
\end{eqnarray}
where
\begin{equation}
\Theta = \frac{c^3\eta}{5 G M}(t_c - t)\;.
\end{equation}
The leading term is just the Newtonian quadrupole phase, Eq.\
(\ref{eq:phi_NQ}).  Each power of $\Theta$ connects to a higher order
in the expansion.  Equation (\ref{eq:2pnPhase}) is taken to ``second
post-Newtonian'' order, meaning that corrections of $(v/c)^4$ are
included.  Corrections to order $(v/c)^6$ are summarized in
{\cite{blanchet06}}.  In addition to the chirp mass ${\cal M}$, the
reduced mass $\mu$ enters $\Phi$ when higher order terms are included.
Including higher pN effects in our wave model makes it possible to
determine both chirp mass and reduced mass, fully constraining the
binary's masses.

Equation (\ref{eq:2pnPhase}) also depends on two parameters, $\beta$
and $\sigma$, which come from the binary's spins and orbit
orientation.  The ``spin-orbit'' parameter $\beta$ is
\begin{equation}
\beta = \frac{1}{2}\sum_{i = 1}^2\left[113\left(\frac{m_i}{M}\right)^2 +
75\eta\right]\frac{\hat{\bf L}\cdot{\bf S}_i}{m_i^2}\;;
\label{eq:beta_def}
\end{equation}
the ``spin-spin'' parameter $\sigma$ is
\begin{equation}
\sigma = \frac{\eta}{48m_1^2m_2^2}\left[721(\hat{\bf L}\cdot{\bf S}_1)
(\hat{\bf L}\cdot{\bf S}_2) - 247{\bf S}_1\cdot{\bf S}_2\right]\;
\label{eq:sigma_def}
\end{equation}
{\cite{bdiww95}}.  These parameters encode valuable information,
especially when spin precession is taken into account.

\subsubsection{Spin precession.}
\label{sec:pn_precession}

\begin{figure}[t]
\begin{center}
\includegraphics[width=4in]{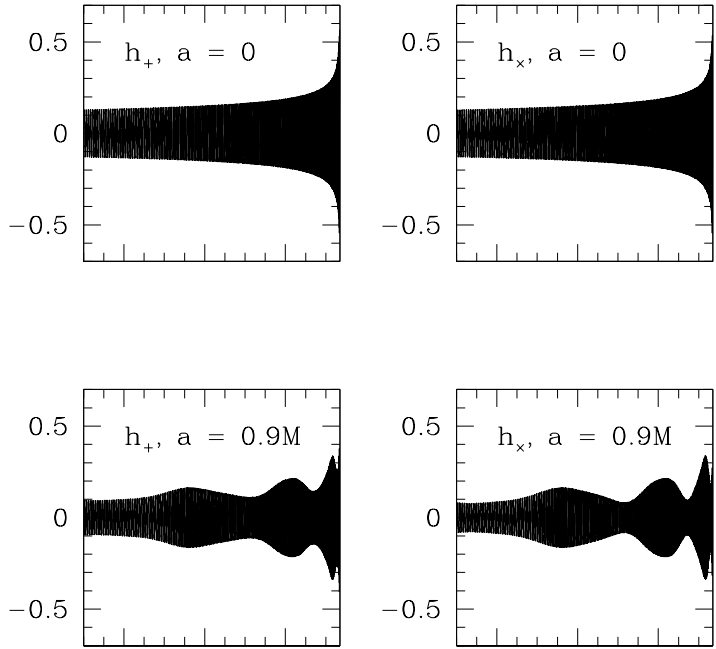}
\end{center}
\vskip -0.5in
\caption{Illustration of precession's impact on a binary's waves.  The
top panels show $h_+$ and $h_\times$ for a binary that contains
nonspinning black holes; the lower panels show the waveforms for a
binary with rapid rapidly rotating ($a = 0.9M$) holes.  The strong
amplitude modulation is readily apparent in this figure.  Less
obvious, but also included, is the frequency modulation that enters
through the spin-dependent orbital phase parameters $\beta$ and
$\sigma$ [cf.\ Eq.\ (\ref{eq:2pnPhase})].}
\label{fig:prec}
\end{figure}

Although the spin vectors ${\bf S}_1$ and ${\bf S}_2$ wiggle around
according to Eqs.\ (\ref{eq:dS1dt}) and (\ref{eq:dS2dt}), the system
must preserve a notion of {\it global} angular momentum.  Neglecting
for a moment the secular evolution of the binary's orbit due to GW
emission, pN encodes the notion that the total angular momentum
\begin{equation}
{\bf J} = {\bf L} + {\bf S}_1 + {\bf S}_2
\end{equation}
must be conserved.  This means ${\bf L}$ must oscillate to compensate
for the spins' dynamics, and guarantees that, when spin precession is
accounted for in our evolutionary models, the phase parameters $\beta$
and $\sigma$ become time varying.  Likewise, the inclination angle
$\iota$ varies with time.  Precession thus leads to phase and
amplitude modulation of a source's GWs.  Figure {\ref{fig:prec}}
illustrates precession's impact, showing the late inspiral waves for
binaries that are identical aside from spin.

Careful analysis shows that accounting for these effects in our wave
model makes it possible to measure the spins of a binary's members, in
many cases with excellent precision {\cite{lh06}}.  By measuring both
masses and spins, instruments such as LISA for example become tools
for tracking the cosmic evolution of black hole masses and spins,
opening a window onto the growth of these objects from early
cosmological epochs.

\section{Extreme mass ratio binary waves}
\label{sec:emri}

We conclude by examining waves from extreme mass ratio binaries ---
stellar mass (roughly $1-100\,M_\odot$) compact bodies spiraling into
a much more massive (roughly $10^5 - 10^7\,M_\odot$) black holes.
Such systems are very well modeled using black hole perturbation
theory, so we begin with a quick review of this subject.

\subsection{Brief overview of black hole perturbation theory}
\label{sec:perturb_formalism}

Black hole perturbation theory can be developed much like the weak
gravity limit described in Sec.\ {\ref{sec:waveform}}, replacing the
flat spacetime metric $\eta_{\alpha\beta}$ with the spacetime of a
black hole:
\begin{equation}
g_{\mu\nu} = g_{\mu\nu}^{\rm BH} + h_{\mu\nu}\;.
\end{equation}
For astrophysical scenarios, one uses the Schwarzschild (non-rotating
black hole) or Kerr (rotating) solutions for $g_{\mu\nu}^{\rm BH}$.
It is straightforward (though somewhat tedious) to then develop the
Einstein tensor for this spacetime, keeping terms only to first order
in the perturbation $h$.

This approach works very well when the background is non-rotating,
\begin{equation}
ds^2 = g_{\mu\nu}^{\rm BH} dx^\mu dx^\nu = -\left(1 -
\frac{2\hat M}{r}\right)dt^2 + \frac{dr^2}{\left(1 - 2\hat M/r\right)} +
r^2d\Omega^2\;,
\end{equation}
where $d\Omega^2 = d\theta^2 + \sin^2\theta d\phi^2$ and $\hat M =
GM/c^2$.  Our discussion for this special case is adapted from
{\cite{rezzolla03}}.  Because the background is spherically symmetric,
we decompose the perturbation into spherical harmonics.  For example,
under rotations in $\theta$ and $\phi$, $h_{00}$ should transform as a
scalar.  We thus put
\begin{equation}
h_{00} = \sum_{lm} a_{lm}(t,r) Y_{lm}(\theta,\phi)\;.
\end{equation}
The components $h_{0i}$ transform like components of a 3-vector, and
can be expanded in vector harmonics; $h_{ij}$ can be expanded in
tensor harmonics.  One can decompose further with parity: Even
harmonics acquire a factor $(-1)^l$ when $(\theta,\phi) \to (\pi -
\theta, \phi + \pi)$; odd harmonics acquire a factor $(-1)^{l+1}$.

By imposing these decompositions, choosing a particular gauge, and
requiring that the spacetime satisfy the vacuum Einstein equation
$G_{\mu\nu} = 0$, we find an equation that governs the perturbations.
Somewhat remarkably, the $t$ and $r$ dependence for all components of
$h_{\mu\nu}$ for given spherical harmonic indices $(l,m)$ can be
constructed from a function $Q(t,r)$ governed by the simple equation
\begin{equation}
\frac{\partial^2 Q}{\partial t^2} - \frac{\partial^2 Q}{\partial
r_*^2} - V(r)Q = 0\;,
\label{eq:schwarz_pert}
\end{equation}
where $r_* = r + 2 \hat M \ln(r/2\hat M - 1)$.  The potential $V(r)$
depends on whether we consider even or odd parity:
\begin{equation}
V_{\rm even}(r) = \left(1 - \frac{2\hat M}{r}\right) \left[\frac{2q(q +
    1)r^3 + 6q^2\hat M r^2 + 18 q\hat M^2 r + 18\hat M^3} {r^3\left(qr
    + 3\hat M\right)^2}\right]\;,
\label{eq:zerilli_pot}
\end{equation}
where $q = (l - 1)(l + 2)/2$; and
\begin{equation}
V_{\rm odd}(r) = \left(1 - \frac{2\hat M}{r}\right)
\left[\frac{l(l+1)}{r^2} - \frac{6\hat M}{r^3}\right]\;.
\label{eq:reggewheeler_pot}
\end{equation}
For even parity, Eq.\ (\ref{eq:schwarz_pert}) is the {\it Zerilli
equation} {\cite{zerilli70}}; for odd, it is the {\it Regge-Wheeler
equation} {\cite{rw57}}.  See {\cite{rezzolla03}} for further
discussion, including how gauge is chosen and how to construct
$h_{\mu\nu}$ from $Q$.  When the spacetime perturbation is due to a
body orbiting the black hole, these equations acquire a source term.
One can find the waves from an orbiting body by using the source-free
equation to build a Green's function, and then integrating over the
source.

How does this procedure fare for rotating holes?  The background
spacetime,
\begin{eqnarray}
ds^2 = -\left(1 - \frac{2\hat Mr}{\rho^2}\right)dt^2 -
\frac{4 a\hat M r\sin^2\theta}{\rho^2}dt d\phi +
\frac{\rho^2}{\Delta}dr^2 + \rho^2d\theta^2 + \left(r^2 + a^2 +
\frac{2\hat Mr a^2\sin^2\theta}{\rho^2}\right)d\phi^2\;,
\nonumber\\
\label{eq:kerr_metric}
\end{eqnarray}
where
\begin{equation}
a = \frac{|\vec S|}{c M}\;,\qquad
\rho^2 = r^2 + a^2\cos^2\theta\;,\qquad
\Delta = r^2 - 2\hat M r + a^2\;,
\end{equation}
is now nonspherical, and the decomposition into spherical harmonics is
not useful.  One could in principle simply expand $G_{\mu\nu} = 0$ to
first order in $h_{\mu\nu}$ and obtain a partial differential equation
in $t$, $r$, and $\theta$.  (The metric is axially symmetric, so we
can easily separate the $\phi$ dependence.)

Rather than expanding the metric, Teukolsky {\cite{teuk73}} examined
perturbations of curvature:
\begin{equation}
R_{\alpha\mu\beta\nu} =
R^{\rm BH}_{\alpha\mu\beta\nu} +
\delta R_{\alpha\mu\beta\nu}\;.
\end{equation}
The curvature tensor is invariant to first-order gauge
transformations, an attractive feature.  This tensor obeys a nonlinear
wave equation which can be derived from the Bianchi identity; see
{\cite{araa}} for discussion.  By expanding this wave equation to
linear order in $\delta R_{\alpha\mu\beta\nu}$, Teukolsky showed that
perturbations to Kerr black holes are governed by the equation
\begin{eqnarray}
&&
\!\!\!\!\!\!
\left[\frac{(r^2 + a^2)^2 }{\Delta} - a^2\sin^2\theta\right]
\partial^2_{t}\Psi - 4\left[r + ia\cos\theta - \frac{\hat M(r^2 -
a^2)}{\Delta}\right]\partial_t\Psi
+\frac{4i \hat M a m r}{\Delta}\partial_t\Psi -
\Delta^{2}\partial_r\left(\Delta^{-1}\partial_r\Psi\right)
\nonumber\\
&&
- \frac{1}{\sin\theta}\partial_\theta
\left(\sin\theta\partial_\theta\Psi\right)
- \left[\frac{a^2}{\Delta} - \frac{1}{\sin^2\theta}\right]m^2 \Psi +
4im \left[\frac{a (r - \hat M)}{\Delta} + \frac{i
    \cos\theta}{\sin^2\theta} \right]\Psi - \left(4\cot^2\theta +
2\right) \Psi = {\cal T}\;.  \nonumber\\
\label{eq:teukolsky}
\end{eqnarray}
The field $\Psi$ is a complex quantity built from a combination of
components of $\delta R_{\alpha\mu\beta\nu}$.  It describes a
spacetime's radiation; see {\cite{teuk73}} for details.  (We have
assumed $\Psi \propto e^{im\phi}$.)  Likewise, ${\cal T}$ describes a
source function built from the stress-energy tensor describing a small
body orbiting the black hole.

Somewhat amazingly, Eq.\ (\ref{eq:teukolsky}) separates: putting
\begin{equation}
\Psi = \int d\omega \sum_{lm}
R_{lm\omega}(r)S_{lm\omega}(\theta)e^{im\phi - i\omega t}
\label{eq:teuk_decomp}
\end{equation}
and applying a similar decomposition to the source ${\cal T}$, we find
that $S_{lm\omega}(\theta)$ is a ``spin-weighted spheroidal harmonic''
(a basis for tensor functions in a non-spherical background), and that
$R_{lm\omega}(r)$ is governed by a simple ordinary differential
equation.  $\Psi$ characterizes Kerr perturbations in much the same
way that $Q$ [cf.\ Eq.\ (\ref{eq:schwarz_pert})] characterizes them
for Schwarzschild.  Although the perturbation equations are often
solved numerically, analytic solutions are known \cite{mst96}, and can
dramatically improve one's scheme for solving for black hole
perturbations; see Refs.\ {\cite{ft04,ft05}}.

How do we describe the motion of a small body about a black hole?  The
most rigorous approach is to enforce $\nabla^\mu T_{\mu\nu} = 0$,
where $T_{\mu\nu}$ describes the small body in the spacetime of the
large black hole.  Neglecting the small body's perturbation to the
spacetime, we find the geodesic equation $u^\mu \nabla_\mu u^\nu = 0$,
where $u^\mu$ is the small body's 4-velocity.  Geodesic black hole
orbits have been studied extensively; see, for example, Ref.\
{\cite{mtw}}, Chapter 33.  They are characterized (up to initial
conditions) by three conserved constants: energy $E$, axial angular
momentum $L_z$, and ``Carter's constant'' $Q$.  If the black hole does
not rotate, Carter's constant is related to the orbit's total angular
momentum: $Q(a = 0) = {\bf L}\cdot{\bf L} - L_z^2$.  When the black
hole rotates rapidly, $Q$ is not so easy to interpret; the idea that
it is essentially the rest of the orbit's angular momentum can be
useful.

Taking into account perturbations from the small body, $\nabla^\mu
T_{\mu\nu} = 0$ now implies that the small body follows a ``forced''
geodesic,
\begin{equation}
u^\mu \hat\nabla_\mu u^\nu = f^\nu\;,
\label{eq:selfforce_eom}
\end{equation}
where $\hat\nabla_\mu$ is the covariant derivative in the background
spacetime.  The novel feature of Eq.\ (\ref{eq:selfforce_eom}) is the
{\it self force} $f^\nu$, a correction to the motion of order the
small body's spacetime perturbation.  The self force is so named
because it arises from the body's interaction with its own spacetime
correction.

Computing the gravitational self force near a black hole is an active
area of current research.  It is useful to break the self force into a
{\it dissipative} piece, $f^\nu_{\rm diss}$, which is asymmetric under
time reversal, and a {\it conservative} piece, $f^\nu_{\rm cons}$,
which is symmetric.  Dissipation causes the ``conserved'' quantities
$(E, L_z, Q)$ to decay, driving inspiral of the small body.  Quinn and
Wald {\cite{qw99}} have shown that the rate at which $E$ and $L_z$
change due to $f^\nu_{\rm diss}$ is identical to what is found when
one computes the fluxes of energy and angular momentum encoded by the
Isaacson tensor (\ref{eq:isaacson_tmunu}).

The conservative self force does not cause orbit decay.  ``Conserved''
constants remain conserved, but the orbits are shifted from the
background geodesics.  This reflects the fact that, even neglecting
dissipation, the small body's motion is determined by the full
spacetime, not just the background black hole.  Conservative effects
shift the orbital frequencies by an amount
\begin{equation}
\delta\Omega_x \sim \Omega_x \times (\mu/M)
\end{equation}
[where $x \in (\phi,\theta,r)$].  Because the GWs have spectral
support at harmonics of the orbital frequencies, these small but
non-negligible frequency shifts are directly encoded in the waves that
the binary generates.  Good discussion and a toy model can be found in
{\cite{ppn05}}.

There has been enormous progress in understanding self forces on
orbits around non-rotating holes.  Barack and Sago {\cite{bs07}} have
completed an analysis of the full self force for circular orbits about
a Schwarzschild black hole; generalization to eccentric orbits is in
progress (L.\ Barack, private communication).  An independent approach
developed by Detweiler {\cite{det08}} has been found to agree with
Barack and Sago extremely well; see {\cite{sbd08}} for detailed
discussion of this comparison.

\subsection{Gravitational waves from extreme mass ratio binaries}
\label{sec:emri_waves}

We now discuss the properties of GWs and GW sources as calculated
using perturbation theory.  Our goal is to highlight features of the
Kerr inspiral waveform.  We will neglect the conservative self force,
which is not yet understood for the Kerr case well enough to be
applied to these waves.  When conservative effects are neglected, the
binary can be regarded as evolving through a sequence of geodesics,
with the sequence determined by the rates at which GWs change the
``constants'' $E$, $L_z$, and $Q$.  Modeling compact binaries in this
limit takes three ingredients: First, a description of black hole
orbits; second, an algorithm to compute GWs from the orbits, and to
infer how the waves' backreaction evolves us from orbit to orbit; and
third, a method to integrate along the orbital sequence to build the
full waveform.  A description of this method is given in
{\cite{hetal05}}; we summarize the main results of these three
ingredients here.

\subsubsection{Black hole orbits.}
\label{sec:bhorbits}

\begin{figure}[t]
\begin{center}
\includegraphics[width=5in]{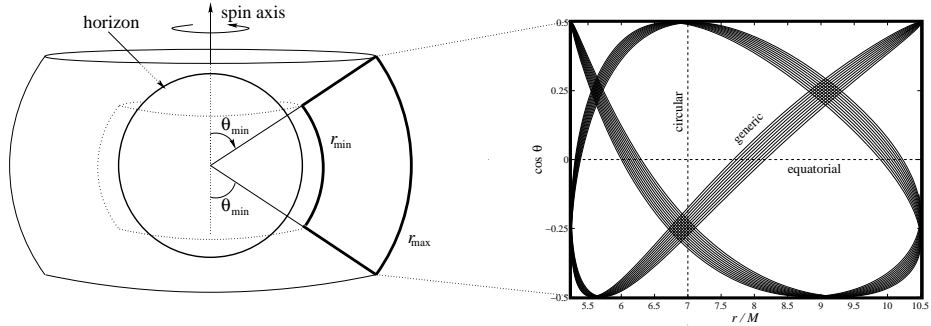}
\end{center}
\vskip -0.25in
\caption{The geometry of a generic Kerr black hole orbit [taken from
{\cite{dh06}}].  This orbit is about a black hole with spin parameter
$a = 0.998M$ (recall $a \le M$, so this represents a nearly maximally
spinning black hole).  The range of its radial motion is determined by
$p = 7GM/c^2$ ($G$ and $c$ are set to 1 in the figure) and $e = 1/3$;
$\theta$ ranges from $60^\circ$ to $120^\circ$.  The left panel shows
the torus in coordinate space this torus occupies.  The right panel
illustrates how a generic orbit ergodically fills this torus.}
\label{fig:torus}
\end{figure}

\begin{figure}[t]
\begin{center}
\includegraphics[width=3.5in]{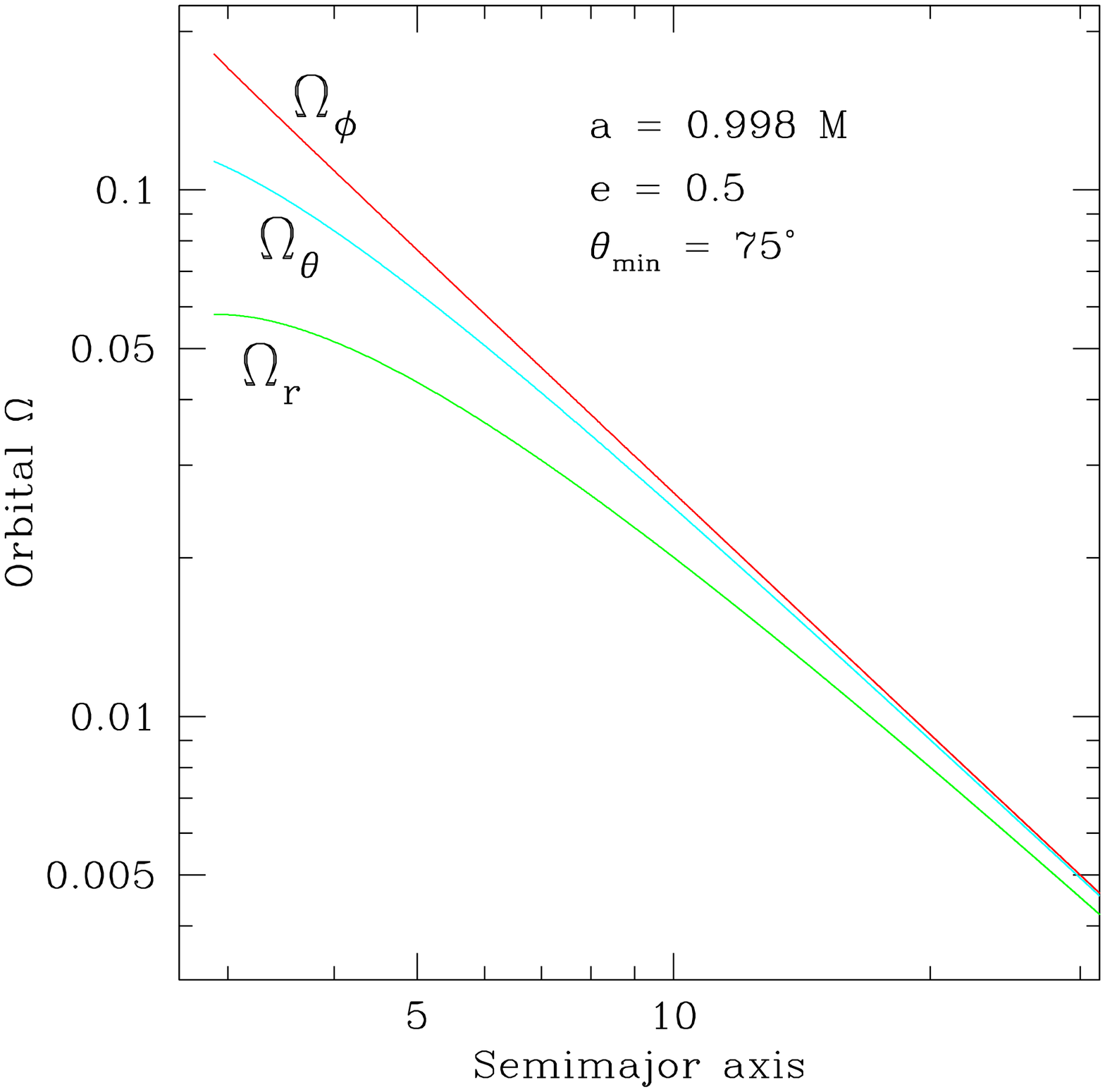}
\end{center}
\caption{Orbital frequencies for generic Kerr black hole orbits.  We
vary the orbits' semilatus rectum $p$, but fix eccentricity $e = 0.5$
and inclination parameter $\theta_{\rm min} = 75^\circ$.  Our results
are plotted as a function of semimajor axis $A = p/\sqrt{1 - e^2}$.
All three frequencies asymptote to the Keplerian value $\Omega =
\sqrt{GM/A^3}$ in the weak field, but differ significantly from each
other in the strong field.}
\label{fig:freqs}
\end{figure}

Motion near a black hole can be conveniently written in the
coordinates of Eq.\ (\ref{eq:kerr_metric}) as $r(t)$, $\theta(t)$, and
$\phi(t)$.  Because $t$ corresponds to time far from the black hole,
this gives a useful description of the motion as measured by distant
observers.  {\it Bound} orbits are confined to a region near the hole.
They have $r_{\rm min} \le r(t) \le r_{\rm max}$ and $\theta_{\rm min}
\le \theta(t) \le \pi - \theta_{\rm min}$, and thus occupy a torus in
the 3-space near the hole's event horizon; an example is shown in
Fig.\ {\ref{fig:torus}}, taken from {\cite{dh06}}.  Selecting the
constants $E$, $L_z$, and $Q$ fully determines $r_{\rm min/max}$ and
$\theta_{\rm min}$.  It is useful for some discussions to
reparameterize the radial motion, defining an eccentricity $e$ and a
semi-latus rectum $p$ via
\begin{equation}
r_{\rm min} = \frac{p}{1 + e}\;,\qquad
r_{\rm max} = \frac{p}{1 - e}\;.
\end{equation}
For many bound black hole orbits, $r(t)$, $\theta(t)$, and $\phi(t)$
are periodic {\cite{schmidt02, dh04}}.  (Exceptions are orbits which
plunge into the hole; we discuss these below.)  Near the hole, the
time to cover the full range of $r$ becomes distinct from the time to
cover the $\theta$ range, which becomes distinct from the time to
cover $2\pi$ radians of azimuth.  One can say that spacetime curvature
splits the Keplerian orbital frequency $\Omega$ into $\Omega_r$,
$\Omega_\theta$, and $\Omega_\phi$.  Figure {\ref{fig:freqs}} shows
these three frequencies, plotted as functions of semi-major axis $A$
for fixed values of $e$ and $\theta_{\rm min}$.  Notice that all three
approach $\Omega \propto A^{-3/2}$ for large $A$.

\subsubsection{Gravitational radiation from orbits.}
\label{sec:orbitwaves}

Because their orbits are periodic, GWs from a body orbiting a black
hole will have support at harmonics of the orbital frequencies.  One
can write the two polarizations
\begin{equation}
h_+ + i h_\times = \sum H_{mkn}
e^{i\omega_{mkn}t}\;,\qquad\mbox{where}
\label{eq:fd_waveform}
\end{equation}
\begin{equation}
\omega_{mkn} = m\Omega_\phi + k\Omega_\theta + n\Omega_r\;.
\label{eq:harmonics}
\end{equation}
The amplitude $H_{mkn}$ can be found by solving the Teukolsky equation
(\ref{eq:teukolsky}) using the decomposition (\ref{eq:teuk_decomp});
details for the general case can be found in {\cite{dh06}}.

The expansion (\ref{eq:fd_waveform}) does not work well for orbits
that plunge into the black hole; those orbits are not periodic, and
cannot be expanded using a set of real frequencies.  A better way to
calculate those waves is to solve the Teukolsky equation
(\ref{eq:teukolsky}) {\it without} introducing the decomposion
(\ref{eq:teuk_decomp}).  Results for waves from plunging orbits in the
language of perturbation theory were first given by Damour, Nagar, and
Tartaglia {\cite{ndt07}}; Sundararajan {\cite{sundararajan08}} has
recently extended the cases that we can model to full generality.

As mentioned above, it is fairly simple to compute the flux of energy
$\dot E$ and angular momentum $\dot L_z$ from the Isaacson tensor,
Eq.\ (\ref{eq:isaacson_tmunu}), once the waves are known.  Recent work
{\cite{ganz07}} has shown that a similar result describes $\dot Q$.
Once $\dot E$, $\dot L_z$, and $\dot Q$ are known, it is
straightforward to evolve the orbital elements $r_{\rm min/max}$ and
$\theta_{\rm min}$, specifying the sequence of orbits through which
GWs drive the system.  Once that sequence is known, it is
straightforward to build the worldline that a small body follows as it
spirals into the black hole.  From the worldline, we can build a
source function ${\cal T}(t)$ for Eq.\ (\ref{eq:teukolsky}) and
compute the evolving inspiral waves.

\subsection{Mapping black hole spacetimes}
\label{sec:bothros}

Extreme mass ratio GW events may allow a unique and powerful
measurement: We may use them to ``map'' the spacetimes of black holes
and test how well they satisfy the stringent requirements of GR.  As
discussed above, an extreme mass ratio inspiral is essentially a
sequence of orbits.  Thanks to the mass ratio, the small body moves
through this sequence slowly, spending a lot of time ``close to'' any
orbit in the sequence.  Also thanks to the mass ratio, each orbit's
properties are mostly determined by the larger body.  In analogy to
{\it geodesy}, the mapping of earth's gravity with satellite orbits,
one can imagine {\it bothrodesy}\footnote{This name was coined by
Sterl Phinney, and comes from the word $\beta o\theta\!\rho
o\varsigma$, which refers to a sacrificial pit in ancient Greek.  This
author offers an apology to speakers of modern Greek.}, the mapping of
a black hole's gravity by studying the orbits of inspiraling
``satellites.''

In more detail, consider first Newtonian gravity.  The exterior
potential of a body of radius $R$ can be expanded in a set of
multipole moments:
\begin{equation}
\Phi_N = -\frac{GM}{r} + G\sum_{l = 2}^\infty
\left(\frac{R}{r}\right)^{l + 1} M_{lm} Y_{lm}(\theta,\phi)\;.
\label{eq:earth_pot}
\end{equation}
Studying orbits allows us to map the potential $\Phi_N$, and thus to
infer the moments $M_{lm}$.  By enforcing Poisson's equation in the
interior, $\nabla^2\Phi_N = 4\pi G\rho$, and then matching at the
surface $R$, one can relate the moments $M_{lm}$ to the distribution
of matter.  In this way, orbits allow us to map in detail the
distribution of matter in a body like the earth.

Bothrodesy applies the same idea to a black hole.  The spacetime of
any stationary, axisymmetric body can be described by a set of ``mass
moments'' $M_l$, similar to the $M_{lm}$ of Eq.\ (\ref{eq:earth_pot});
and a set of ``current moments'' $S_l$ which describe the distribution
of mass-energy's {\it flow}.  The moments of a black hole take a
simple, special form: for a Kerr black hole (\ref{eq:kerr_metric})
with mass $M$ and spin parameter $a$,
\begin{equation}
M_l + i S_l = M(ia)^l\;.
\label{eq:kerr_moments}
\end{equation}
A black hole has a mass moment $M_0 = M$ and a current moment $S_1 =
aM$ (i.e., the magnitude of its spin is $aM$, modulo factors of $G$
and $c$). {\it Once those moments are known, all other moments are
fixed if the Kerr solution describes the spacetime.}  This is a
restatement of the ``no hair'' theorem {\cite{carter71, robinson75}}
that a black hole's properties are set by its mass and spin.

The facts that an object's spacetime and orbits are determined by its
multipoles, and that the Kerr moments take such a simple form,
suggests a consistency test: Develop an algorithm for mapping the
multipolar structure by studying orbits, and check that the $l \ge 2$
moments satisfy Eq.\ (\ref{eq:kerr_moments}).  Ryan \cite{ryan95}
first demonstrated that such a measurement can be done, and Brink
\cite{brink08} has recently clarified what must be done for such
measurements to be done in practice.  Collins and Hughes \cite{ch04}
took the first steps in formulating this question as a null experiment
(with the Schwarzschild solution as the null hypothesis).  Glampedakis
and Babak \cite{gb06} formulated a similar approach appropriate to
Kerr black holes; Vigeland and Hughes {\cite{vh10}} have recently
extended the Collins and Hughes formalism in that direction.

A robust test of the Kerr solution is thus a very likely outcome of
measuring waves from extreme mass ratio captures.  If testing metrics
is not your cup of tea, precision black hole metrology may be: In the
process of mapping a spacetime, one measures with exquisite accuracy
both the mass and the spin of the large black hole.  Barack and Cutler
\cite{bc04} have found that in most cases these events will allow us
to determine both the mass and the spin of the large black hole with
$0.1\%$ errors are better.  GW measurements will give us a precise
picture of these amazing objects.

\section*{Acknowledgments}

Portions of this proceedings article were adapted from previous
reviews I have written or cowritten (Refs.\ {\cite{njp, araa}}).  Much
of the research discussed here was done in collaboration with my
collaborators Steve Drasco, \'Eanna Flanagan, and Gaurav Khanna, as
well as my current and former graduate students Nathan Collins, Ryan
Lang, Pranesh Sundararajan, and Sarah Vigeland.  My group's research
in gravitational waves and compact binaries is supported by NSF Grant
PHY-0449884 and NASA Grant NNX08AL42G; some of the work discussed here
was also supported by NASA Grant NNG05G105G and the MIT Class of 1956
Career Development Fund.  I gratefully acknowledge the support of the
Adam J.\ Burgasser Chair in Astrophysics at MIT in completing this
conference writeup.

\end{document}